\documentclass[letterpaper]{ptephy_v1}

\usepackage{amsmath}
\usepackage{graphics}
\usepackage{url}

\usepackage{color}

\begin{document}

\title{Black hole ringdown echoes and howls}


\author{\name{Hiroyuki Nakano}{1,2},
\name{Norichika Sago}{3}, 
\name{Hideyuki Tagoshi}{4}
and \name{Takahiro Tanaka}{2,5}}
\address{
${}^1$\affil{1}{Faculty of Law, Ryukoku University, Kyoto 612-8577, Japan}
\\
${}^2$\affil{2}{Department of Physics, Kyoto University, Kyoto 606-8502, Japan}
\\
${}^3$\affil{3}{Faculty of Arts and Science, Kyushu University, Fukuoka 819-0395, Japan}
\\
${}^4$\affil{4}{Graduate School of Science, Osaka City University, Osaka 558-8585, Japan}
\\
${}^5$\affil{5}{Yukawa Institute for Theoretical Physics, Kyoto University, Kyoto 606-8502, Japan}
}

\begin{abstract}
Recently the possibility of detecting echoes of ringdown gravitational
waves from binary black hole mergers was shown.
The presence of echoes is expected if the black hole is surrounded
by a mirror that reflects gravitational waves near the horizon.
Here, we present slightly more 
sophisticated templates motivated by a waveform which is obtained
by solving the linear perturbation equation 
around a Kerr black hole with a complete reflecting boundary condition
in the stationary traveling wave approximation.
We estimate that the proposed template
can bring about $10\%$ improvement in the signal-to-noise ratio.  
\end{abstract}

\subjectindex{E31, E02, E01, E38}

\maketitle

\section{Introduction}

Direct gravitational wave (GW) detections
\cite{TheLIGOScientific:2016qqj,Abbott:2016nmj}
give us various opportunities to discuss the nature of black holes (BHs).
Abedi, Dykaar, and Afshordi~\cite{Abedi:2016hgu}
have reported an interesting result by reanalyzing GWs from
two binary black hole (BBH) merger events GW150914 and GW151226,
and a candidate LVT151012 in the first Advanced LIGO
observing run (O1)~\cite{TheLIGOScientific:2016pea}.
In the merger events, 
the possibility of detecting echoes of ringdown GWs from BBH mergers was shown.
The GW echoes are expected
if the BH is surrounded by a mirror that reflects GWs
near the event (or apparent)
horizon~\cite{Cardoso:2016rao,Cardoso:2016oxy},
which is motivated by 
the quantization of the horizon area~\cite{Bekenstein, Mukhanov}
(see also Ref.~\cite{Barcelo:2017lnx} and references therein 
for recent work on various quantum effects). 
There has been comment~\cite{Ashton:2016xff}
on the method of data analysis and significance
estimation used in Ref.~\cite{Abedi:2016hgu},
with a corresponding reply~\cite{Abedi:2017isz}.

In the analysis of Ref.~\cite{Abedi:2016hgu},
a simple template has been used for the analysis of GW echoes
in the matched filtering method
[see a similar template shown in Eq.~\eqref{eq:simpleT}].
In this paper, we consider a Kerr BH~\cite{Kerr:1963ud} with mass $M$
and spin parameter $a$, and calculate echo waveforms
by solving the linear perturbation equation 
with a complete reflecting boundary condition in Sect.~\ref{sec:TRB}.
We introduce waveforms obtained by using
the stationary traveling wave approximation,
which are valid when the interference between adjacent echoes can be ignored.
Here, we also give a brief comment on a problem related to 
the super-radiant instability~\cite{Friedman,Vilenkin:1978uc}.
In Sect.~\ref{sec:temp},
we present slightly more sophisticated templates
[see Eq.~\eqref{waveform} with Eq.~\eqref{eq:fitting}]
based on the echo waveforms.
We find that the reflection rate depends on the frequency
while the rate adopted in Ref.~\cite{Abedi:2016hgu}
is independent of the GW frequency.
Using an inner product of waveforms, we evaluate
{\it decline rates} of the signal amplitude in Sect.~\ref{sec:dec}.
Section~\ref{sec:sum} is devoted to a brief summary.

\section{Totally reflecting boundary for Teukolsky and Sasaki--Nakamura equations}
\label{sec:TRB}

The perturbation equation for GWs on the Kerr background 
is written in a single equation called Teukolsky equation
\cite{Teukolsky:1972my,Teukolsky:1973ha,Press:1973zz,Teukolsky:1974yv}. 
The equation is separable. Assuming that $t$ and $\varphi$ dependencies are 
given by $\propto e^{-i\omega t+im\varphi}$, the
source-free radial equation is given by 
\begin{equation}
 \left[\Delta \frac{d^2}{dr^2}+2(s+1)(r-M)\frac{d}{dr}+
 \left(\frac{K^2-2is(r-M)K}{\Delta}+4ir\omega s -\lambda \right)\right]R_s=0
\,,
\end{equation}
where $\lambda$ is related to the eigenvalue of the angular function, $E$,
as $\lambda:=E-2a\omega m+a^2\omega^2-s(s+1)$, 
\begin{eqnarray}
  K =  (r^2+a^2)\,\omega -a m\,, \quad
  \Delta = r^2-2M r +a^2\,,
\end{eqnarray}
and $s$ can take $\pm 2$. 

In the following analysis, we assume for given $\omega$
that the amplitude and the phase shift of reflected and transmitted waves
are the same as those of stationary traveling waves
when some incident waves collide with the potential,
and that the amplitude of waves reflected at a complete reflecting boundary
is the same as that before the reflection.
Although we use the above stationary traveling wave approximation,
the approximate solution can be considered as the exact one
when each echo is well separated.

The conserved current corresponding to the energy flux and angular momentum 
flux can be read from the Wronskian relation. 
When the potential is real, the complex conjugate of a solution also 
becomes a solution, and hence we can establish a non-trivial 
Wronskian relation between a solution and its complex conjugate. 
In the present case, it is not so straightforward to obtain the 
relation for conserved quantities from the constancy of the Wronskian  
since the coefficients of the equation are complex valued. 
However, if we define a new radial function 
\begin{equation}
{\cal Y}_s:=\Delta^{s/2}(r^2+a^2)^{1/2} R_s\,,
\end{equation}
${\cal Y}_s$ and ${\cal Y}^*_{-s}$ satisfy the same radial equation in 
the form of $\left(d^2/{dr^*}^2-{\cal V}\right){\cal Y}=0$
\cite{Sasaki:1981kj,Sasaki:1981sx,Nakamura:1981kk},
and hence we find that 
\begin{equation}
 (\partial_{r^*}{\cal Y}_s) {\cal Y}^*_{-s}-{\cal Y}_s(\partial_{r^*}{\cal Y}^*_{-s})
\label{Wronskian}
\end{equation}
with $dr^*:=dr\Delta/(r^2+a^2)$, becomes constant in $r$. 

A radial function $R_s$ for the spin $s$ can be transformed into 
a radial function for the spin $-s$ by using the Teukolsky--Starobinsky relations:
\begin{eqnarray}
{\cal D}^4 R_{-2} &=& \frac14 R_2\,, \cr
\left({\cal D}^\dag\right)^4 R_{2} &=& 4|C|^2\Delta^{-2} R_{-2}\,,
\label{STrelations}
\end{eqnarray} 
with ${\cal D} := \partial _{r}-iK/\Delta$ and 
${\cal D^\dag} := \partial _{r}+iK/\Delta$, 
where 
\begin{eqnarray}
|C|^2&=&(Q^2+4a\omega m -4a^2\omega^2)
    \left[(Q-2)^2+36a\omega m -36a^2\omega^2\right]\cr
    &&+48(2Q-1)(2a^2\omega^2-a\omega m)+144\omega^2(M^2-a^2)\,,
\end{eqnarray}
is the Starobinsky--Churilov constant~\cite{Starobinsky:1973aij,Starobinsky:1973aij2}.
Here, $Q:=E+a^2\omega^2-2a\omega m$.

Our focus is on the solution that satisfies the following 
boundary condition:
\begin{eqnarray}
R_2 & = & \left\{
\begin{array}{ll}
 Y_{\rm out} \frac{e^{i\omega r^*}}{r^5}, & \mbox{for}~r^*\to\infty\,,\cr
 Y_{\rm up} \Delta^{-2} e^{-i k r^*}
     +Y_{\rm down} e^{i k r^*}, \qquad &\mbox{for}~r^*\to -\infty\,,
\end{array}\right.\cr
R_{-2} & = & \left\{
\begin{array}{ll}
 Z_{\rm out} r^3 e^{i\omega r^*}, & \mbox{for}~r^*\to\infty\,,\cr
 Z_{\rm up} \Delta^{2} e^{-i k r^*}
     +Z_{\rm down} e^{i k r^*}, \qquad & \mbox{for}~r^*\to -\infty\,, 
\end{array}\right.
\end{eqnarray}
which means that there is no incoming wave from a large radius. 

Then, the relations \eqref{STrelations} give
\begin{eqnarray}
  4\omega^4 Y_{\rm out} & = & C^* Z_{\rm out}\,,\cr
  CY_{\rm up} 
     & = & 64(2Mr_+)^4 k(k+4i\epsilon)(k^2+4\epsilon^2)  Z_{\rm up}\,,\cr
  4(2Mr_+)^4 k(k-4i\epsilon)(k^2+4\epsilon^2) Y_{\rm down} 
     & = & C^* Z_{\rm down}\,,
 \end{eqnarray}
 where $r_+:=M+\sqrt{M^2-a^2}$, which is the outer horizon radius, 
 $k:=\omega-ma/(2Mr_+)$ and 
$\epsilon=\sqrt{M^2-a^2}/(4Mr_+)$. 
On the other hand, the Wronskian relation \eqref{Wronskian} gives 
 \begin{equation}
 (2Mr_+)(ik+4\epsilon) \left(-Y_{\rm up}Z^*_{\rm up}
 +Y_{\rm down}Z^*_{\rm down}\right)=i\omega Y_{\rm out} Z^*_{\rm out}\,.
 \end{equation}
Combining these relations, we obtain 
 \begin{eqnarray}
 |Y_{\rm out}|^2=
 \frac{(2Mr_+)^5 k (k^2+4\epsilon^2)(k^2+16\epsilon^2)}{\omega^5} |Y_{\rm down}|^2
   -\frac{|C|^2 |Y_{\rm up}|^2}{256k\omega^5 (2Mr_+)^3(k^2+4\epsilon^2)}\,,\cr
 |Z_{\rm out}|^2=
  \frac{\omega^3 |Z_{\rm down}|^2}{k (2Mr_+)^3(k^2+4\epsilon^2)}
  -\frac{256 \omega^3 (2Mr_+)^5 k (k^2+4\epsilon^2)(k^2+16\epsilon^2)}{|C|^2}  |Z_{\rm up}|^2
  \,.
 \label{eq:immed}
 \end{eqnarray}
The squared reflection rate by the potential barrier would be defined by the 
ratio of the first and the second terms on the right-hand side as 
\begin{eqnarray}
{\rm R}&=& 
\frac{|C|^2}{256 k^2 (2Mr_+)^8(k^2+4\epsilon^2)^2(k^2+16\epsilon^2)}
  \frac{ |Y_{\rm up}|^2}{|Y_{\rm down}|^2}\nonumber\\
  &=&
  \frac{256 k^2 (2Mr_+)^8(k^2+4\epsilon^2)^2 (k^2+16\epsilon^2)}{|C|^2}
  \frac{ |Z_{\rm up}|^2}{|Z_{\rm down}|^2}
  \,. 
\label{eq:R}
\end{eqnarray}
We immediately notice from Eq.~\eqref{eq:immed}
that the reflection rate $\sqrt{\rm R}$
is less than unity as long as 
$k/\omega$ is positive, which is expected in ordinary scattering problems.
However, this condition is violated in the super-radiant frequency band 
specified by $0\lessgtr\omega\lessgtr m\omega_+:=ma/(2Mr_+)$ for $m\gtrless 0$.

The phase shift at the reflection by the potential barrier might be
quantified by 
\begin{eqnarray}
 \phi_2(f) &:=& \arg [Y_{\rm up}/Y_{\rm down}]\,, \cr
 \phi_{-2}(f) &:=& \arg [Z_{\rm up}/Z_{\rm down}] \,.
\end{eqnarray}
To predict how the waveform of the echoes is modified at each bounce, 
we also need the phase shift at the boundary near the horizon, which is 
highly model dependent. Therefore, here we use $\phi_{-2}(f)$ just 
for the purpose of order of magnitude estimation of the frequency 
dependence of the phase shift.

\begin{figure}[!t]  
\begin{center}
\includegraphics[bb=0 0 750 497, width=0.7\textwidth,clip=true]{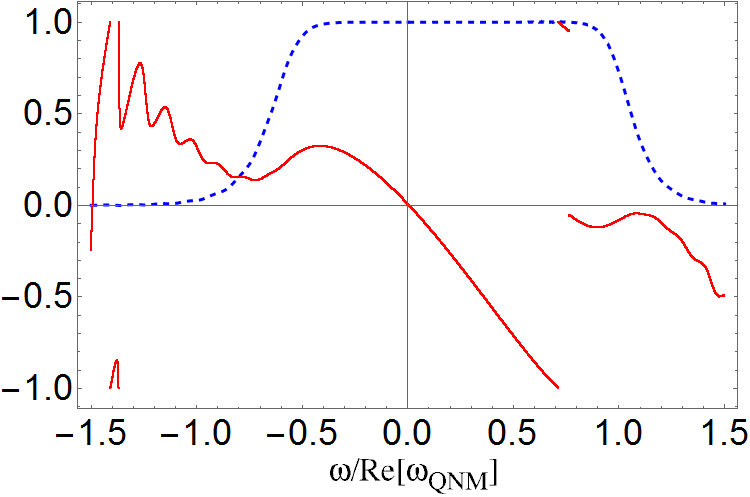}
\end{center}
 \caption{$\sqrt{\rm R}$ (blue dashed curve)
and $(\phi_{-2} \mod 2\pi)/\pi$ (red solid curve)
for $a/M \equiv q=0.7$ and $m=2$.
The horizontal axis is $\omega$ normalized by $\Re(\omega_{\rm QNM})$.}
\label{Fig:example}
\end{figure}  

In Fig.~\ref{Fig:example},
$\sqrt{\rm R}$ (blue solid curve) and $\phi_{-2}$ (red dashed curve) for 
$a/M \equiv q=0.7$ and $m=2$ are displayed. 
The excess of $\sqrt{\rm R}$ above unity due to super-radiance
is at most about $0.005$ 
at around $f\approx 0.75\Re(\omega_{\rm QNM})/2\pi$, which is invisible 
in this plot. Here, $\omega_{\rm QNM}$ represents the least damped quasinormal mode 
frequency for the $m=2$ mode of the usual Kerr spacetime.
Due to the super-radiant instability, GW modes are exponentially amplified
as the wave bounces back and forth 
between the boundary near the horizon and the potential barrier.
The super-radiant modes will survive
as long as rotational and time-translation symmetries are maintained,
and this super-radiant amplification is unavoidable if 
a complete reflecting boundary condition is imposed
around the horizon of a Kerr BH.

The super-radiant amplification looks dangerous.
There are extensive works on this problem
(see, e.g., Refs.~\cite{Cardoso:2007az,Cardoso:2008kj,Pani:2010jz}
and Ref.~\cite{Brito:2015oca} for a review).
The latest analysis~\cite{Maggio:2017ivp} shows that
the time scale can be larger than the age of the Universe
if the location of the reflection boundary is sufficiently far from
the horizon.
The above means that if BHs have a complete reflecting boundary
at a distance of the order of the Planck length from the horizon,
all astrophysical BHs become non-rotating, 
i.e., Schwarzschild BHs.
If we observe GW howls due to the super-radiant amplification,
it means that only Schwarzschild BHs can exist in our universe.

The above consideration
works only in the linear perturbation.
When the super-radiant mode grows nonlinearly,
the axisymmetry of the background spacetime will be broken,
and the super-radiance will be saturated due to the mode-mixing of $m$ modes.
Using the amplification factor $0.005$ 
at around $f\approx 0.75\Re(\omega_{\rm QNM})/2\pi$
and the interval between echoes
$\Delta t_{\rm echo} = 8M \ln (M/\ell_{p}) + O(q^2)$~\cite{Abedi:2016hgu},
where $\ell_{p}$ is the Planck length,
the growing time scale of the super-radiance is roughly estimated as
$T_{\rm SR} \sim 50$\,s for $M=60 M_\odot$. 
Assuming that the super-radiant mode saturates
due to the nonlinear mode-mixing and
the peak amplitude of comparable-mass BBH mergers is $h \sim 10^{-1} (M/D)$,
the saturated amplitude
is roughly obtained from
the ratio between the growing time scale of the super-radiance
and the oscillating time scale of the wave,
$1/(\Re(\omega_{\rm QNM})T_{\rm SR}) \sim 10^{-4}$,
and the strain will be $h \sim 10^{-4} (M/D)$
where $D$ denotes the distance between the source and an observer.
This is three orders of magnitude lower than the peak amplitude of the BBH merger.

On the other hand, in the current GW observations,
BH spin has been observed for one event, GW151226.
According to Ref.~\cite{Abbott:2016nmj},
at least one of the BHs seems to have spin greater than $q=0.2$.
Also, there are X-ray observations that suggest the existence of 
BH spin~\cite{Reynolds:2013qqa}.  
In order not to contradict these observations,
it would be natural to assume
that the super-radiant instability is suppressed by some unknown mechanism,
for example a non-vanishing absorption of the mirror~\cite{Maggio:2017ivp}
(see also Ref.~\cite{Price:2017cjr} for a careful treatment of the boundary).

Also, if the BH's horizon is replaced by
a complete reflecting boundary, the merger process of two bodies
might be different from the usual one derived by general relativity (GR).
Here we note that the remnant BH's mass and spin
have yet to be determined only by the ringdown GW,
and the data analysis has suggested that the remnant mass is smaller
than that predicted by GR from the data (see the top panel of Fig.~4
in Ref.~\cite{TheLIGOScientific:2016src}).
This may be because the gravitational energy cannot be absorbed
to increase the horizon area in the merger.

\begin{figure}[!t]  
\begin{center}
\includegraphics[bb=0 0 750 490, width=0.49\textwidth,clip=true]{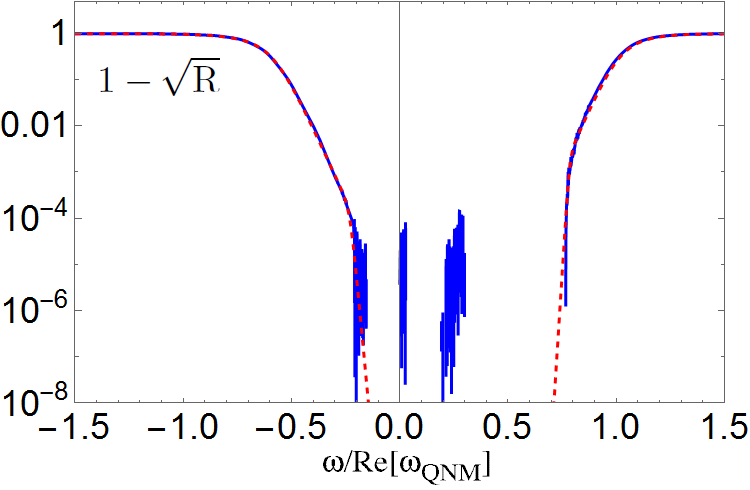}
\includegraphics[bb=0 0 750 500, width=0.49\textwidth,clip=true]{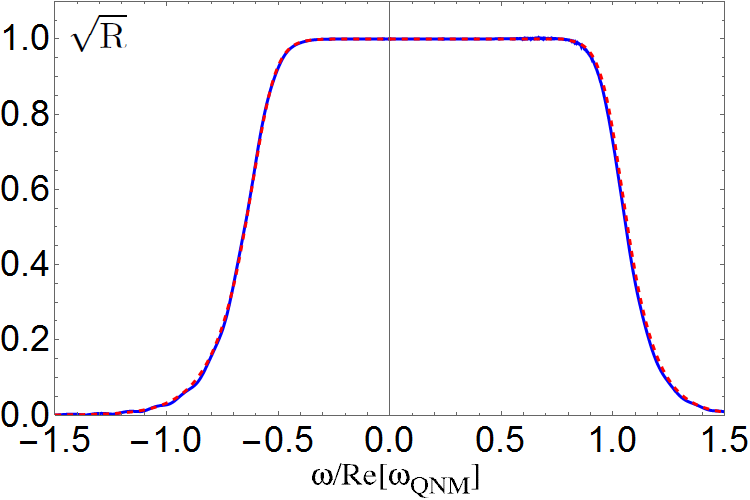}
\end{center}
 \caption{$1-\sqrt{\rm R}$ (left, log plot) and $\sqrt{\rm R}$ (right, linear plot).
Here, we present the case with $q=0.7$.
The blue solid and red dashed curves denote 
the results calculated by Eq.~\eqref{eq:R}
and the fitting functions in Eq.~\eqref{eq:fitting}, respectively.
The horizontal axis is $\omega$ normalized by $\Re(\omega_{\rm QNM})$.}
\label{Fig:fitting}
\end{figure}

Ignoring the super-radiant excess beyond unity
in the reflection rate,
$\sqrt{\rm R}$ can be fitted by the following simple function 
\begin{equation}
 \sqrt{\rm R}\approx 
 \left\{
  \begin{array}{l}
         \displaystyle{\frac{1+e^{-300(x+0.27-q)}+e^{-28(x-0.125-0.6q)}}
       {1+e^{-300(x+0.27-q)}+e^{-28(x-0.125-0.6q)}+e^{19(x-0.3-0.35q)}}} \,, 
        \qquad \mbox{for}~f>0\,,\cr
\cr
        \displaystyle{\frac{1+e^{-200(|x|-0.22+0.1 q)}+e^{-28(|x|-0.39+0.1q)}}
       {1+e^{-200(|x|-0.22+0.1 q)}+e^{-28(|x|-0.39+0.1q)}+e^{16(|x|-0.383+0.09q)}}}\,, 
        \qquad \mbox{for}~f<0\,,
 \end{array}\right.\,
 \label{eq:fitting}
 \end{equation}
where $x:= 2\pi M f=M\omega$. The above fitting formula gives
a good approximation in the range 
$0.6< q <0.8$ of our interest as demonstrated in Fig.~\ref{Fig:fitting}, 
where the dashed lines are the fitting curves.

As we do not know the phase shift caused by the reflection near the horizon,
the overall phase shift cannot be predicted explicitly
without specifying the model for the near-horizon boundary.
Nevertheless, from the plot in Fig.~\ref{Fig:example},
we find that the frequency dependence
of the phase shift from the reflection at the potential barrier is small
if we focus on a narrow frequency band,
say, $0.8< \omega/\Re(\omega_{\rm QNM})<1.2$.
Thus, we assume that the overall phase shift can be approximated well
by a linear function as long as we focus on the frequency band mentioned above. 
Recall that the effect of the phase being linear in $f$ results
in just a shift of the origin of time. Thus, this model parameter  
in the phase shift 
degenerates with the one that determines the 
interval of the arrival times of the echoes. 
As for the waveform of echoes, we expect that 
the least damped quasinormal mode will be most efficiently excited during the 
merger process. Therefore, the spectrum will have a peak around 
$f\approx \Re(\omega_{\rm QNM})/2\pi$. 
This assumption of narrow band 
will justify the approximation of the phase shift by a linear function.

\section{Templates}\label{sec:temp}

We assume that the waveform of the incident wave propagating outward from the boundary 
near the horizon after the merger is provided. 
We denote by $\tilde h(f)$ the hypothetical waveform in Fourier 
space that is obtained if this incident wave passes through the potential barrier 
without any reflection. 
The inverse Fourier transform of $\tilde h(f)$ is not restricted to be real. 
Instead, the corresponding complex time-domain waveform $h(t)$ 
represents $h_+(t)+i h_\times (t)$. 
In reality, the wave experiences a phase shift when reflected at the potential
barrier and also at the boundary near the horizon, which is denoted by $\phi(f)$.  
Using $\tilde h(f)$ and $\phi (f)$, the waveforms of the $n$th 
echoes will be given by 
\begin{equation}
  \tilde h_n(f)=\exp[-i(2\pi f\Delta t+\phi(f))(n-1)](\sqrt{{\rm R}(f)})^{n-1}
\sqrt{1-{\rm R}(f)}\, \tilde h(f)\,,
\label{waveform}
\end{equation}
with the aid of the reflection rate $\sqrt{\rm R}$. 

For $\tilde h(f)$, the following time-domain waveform
(see Eq.~(8) in their paper) was adopted in Ref.~\cite{Abedi:2016hgu} as
\begin{equation}
{\cal M}_{T,I} = \Theta(t,t_0) {\cal M}_I(t) \,,
\end{equation}
where $\Theta$ and ${\cal M}_I$ denote
a smooth cut-off function and the theoretical best-fit waveform
for the BBH merger,
$t_0$ is a free parameter,
and also they assumed that ${\rm R}(f)$ is independent of $f$ and $\phi(f)=0$.  
Here, we propose two simple alternatives for $\tilde h(f)$, which are 
given by 
\begin{equation}
h(t) \propto \frac{e^{-i\omega_{\rm QNM} \tilde t}}
{1+\exp[- 2 \beta t |\Im(\omega_{\rm QNM})|]}\,,
\label{temp2}
\end{equation}
with 
\begin{equation}
  \tilde t=\int^t \frac{dt}{1+\exp[- 2 \alpha t |\Im(\omega_{\rm QNM})|]}\,,
\end{equation} 
and
\begin{equation}
\bar h(t) \propto \frac{e^{-i\omega_{\rm QNM} t}}{1+\exp[-2 \bar\alpha t |\Im(\omega_{\rm QNM})|]}\,,
\label{temp1}
\end{equation}
where $\alpha$ and $\bar\alpha$ are $O(1)$ model parameters, which control the 
smoothness of the onset of the quasinormal mode excitation. 
The parameter $\beta\ll 1$ in the former model is introduced just to 
eliminate the high frequency tail in 
the Fourier transform of the waveform caused by the 
discrete jump between the initial time and final time. 
The first template given in Eq.~\eqref{temp2} phenomenologically 
takes it into account the expectation that the frequencies of the precursor 
of the quasinormal mode ringdown are relatively lower.  

\begin{figure}[!t]  
\begin{center}
\includegraphics[bb=0 0 375 250, width=0.32\textwidth,clip=true]{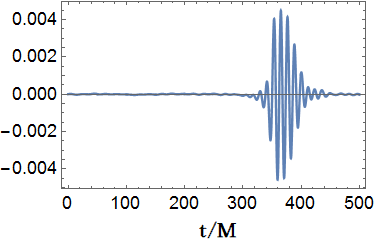}
\includegraphics[bb=0 0 375 250, width=0.32\textwidth,clip=true]{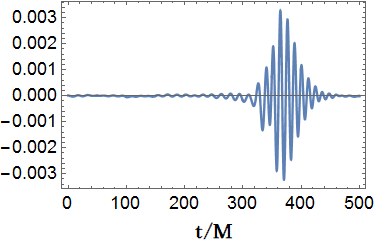}
\includegraphics[bb=0 0 375 250, width=0.32\textwidth,clip=true]{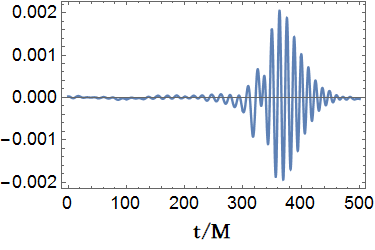}
\includegraphics[bb=0 0 375 250, width=0.32\textwidth,clip=true]{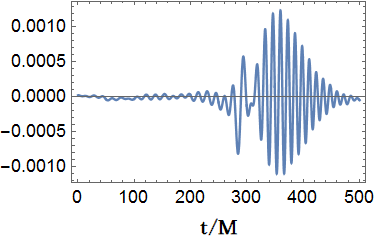}
\includegraphics[bb=0 0 375 250, width=0.32\textwidth,clip=true]{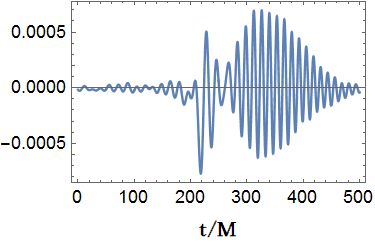}
\end{center}
 \caption{Example of the waveforms of echoes as a function of time.
 From the top left to bottom right,
 the 1st, 2nd, 4th, 8th, and 16th echoes are presented.}
\label{Fig:waveforms}
\end{figure}  

In Fig.~\ref{Fig:waveforms}, we show an example of the waveforms of echoes. 
Here, we used $h$ given in Eq.~\eqref{temp2} with $\alpha=1$ and $\beta=0.1$ 
as the incident waveform, and $q$ is set to 0.7. 
The waveform is significantly modified at each bounce.  
The higher frequency components go through the 
potential barrier easily, and as a result they are quickly lost. 
As the remaining lower frequency modes do not escape so easily, 
the change of the amplitude soon slows down as 
shown in Fig.~4.  
However, even at this stage the waveform continues to be deformed 
because of the frequency-dependent phase shift.  

\begin{figure}[!t]  
\begin{center}
\includegraphics[width=0.8\textwidth]{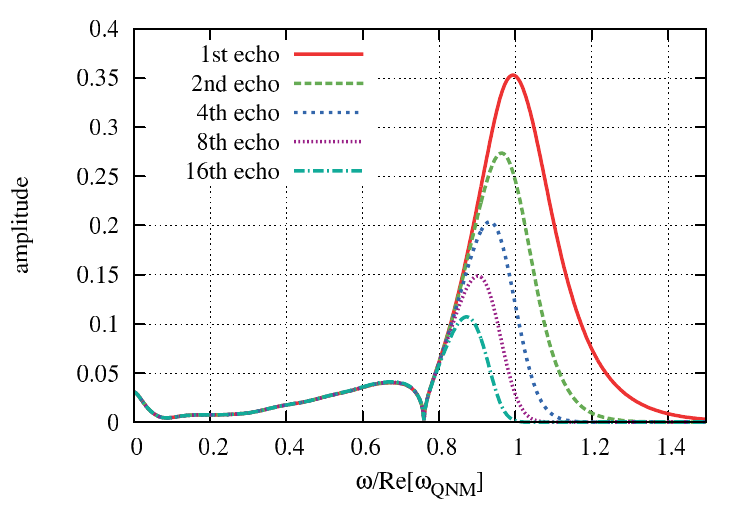}
\end{center}
 \caption{Amplitude of echoes.
The horizontal axis is $\omega$ normalized by $\Re(\omega_{\rm QNM})$.
We show the 1st (largest amplitude), 2nd, 4th, 8th, and 16th (smallest one) echoes
here. Although we observe the tiny super-radiant excess
in the lower frequency region, we ignore it to prepare the templates for echoes.}
\label{Fig:spectrum}
\end{figure}  

\section{Decline rate}\label{sec:dec}

From the waveform in Eq.~\eqref{waveform}, one may naively think that 
the amplitude of the $n$th echo decays like $(\sqrt{\rm R})^n$. 
However, ${\rm R}$ is a function of $f$. After a few bounces the 
echoes should be dominated by the transition frequency region, 
where ${\rm R}$ varies from 0 to unity. Therefore, it is not so straightforward 
to estimate the expected decline rate of the echoes. 

In this section, assuming Eq.~\eqref{temp2} with Eq.~\eqref{waveform}
to be the real signal,
we evaluate the decline rates
which are denoted as $A_n$, $B_n$ and $C_n$,
by using three different templates: 
Eq.~\eqref{temp2} with Eq.~\eqref{waveform},
Eq.~\eqref{temp1} with Eq.~\eqref{waveform},
and Eq.~\eqref{eq:simpleT} below.

First, we calculate 
\begin{equation}
 A_n={(h_n|h_n)}\,, 
\end{equation}
where the inner product is simply defined by
\begin{equation}
  (h|g):= \left|\int df \tilde h(f) \tilde g^*(f) \right| \,,  
\end{equation}
instead of referring to a specific noise curve since 
we do not wish to fix the mass of the BH;
this can be justified since we are focusing on a narrow frequency
band around $\omega_{\rm QNM}$.
The slowest decline given by $A_n$ is realized only when we know 
the exact waveform of the echoes. 

In reality, we can at most give a naive guess for the waveform of the echoes. 
To simulate such a situation,
as a template waveform, we adopt an alternative guess of the waveform $\bar h_n$
given in Eq.~\eqref{temp1} with $\bar\alpha=5$
and with the same quasinormal mode frequency as the incident waveform.
This corresponds to using the remnant BH parameters
derived from data analysis of a usual BBH merger.
The phase shift at each bounce is approximated with a linear function
$\phi(f)=\phi_0+\phi_1 f$. $\phi_1$ can be absorbed by 
the parameter $\Delta t$. Hence, this waveform has two parameters to determine 
from fitting with the data.  
With this choice, 
we evaluate 
\begin{equation}
 B_n=(h_n|{\bar h}_n)\,, 
\end{equation}
marginalizing  $\phi_0$ and $\Delta t$ to maximize 
\begin{equation}
 \rho_B=\sum_n(h_n|{\bar h}_n)\left/
      \sqrt{\left(\sum_n(h_n|h_n)\right)\left(\sum_n(\bar h_n|{\bar h}_n)\right)}\right.\,. 
\end{equation}

For comparison, we consider 
\begin{equation}
 C_n=(h_n|{\bar h}'_n)\,,
\end{equation}
with
\begin{equation}
{\bar h}'_n(t)=\gamma^n \bar h_1(t-(n-1)\Delta t)\,,
\label{eq:simpleT}
\end{equation}
which may mimic the template used in the analysis in 
Ref.~\cite{Abedi:2016hgu}.
We may introduce $\phi_{-2}(f)$ shown in Fig.~\ref{Fig:example}
which holds for the Dirichlet boundary condition
(not for the Teukolsky equation but for the Sasaki--Nakamura equation)
used in Ref.~\cite{Abedi:2016hgu}
because of no relative phase shift between different frequencies
added at the reflection on the boundary.
In this template the parameters 
$\gamma$ and $\Delta t$ are marginalized to maximize 
 \begin{equation}
 \rho_{C}=\sum_n(h_n|{\bar h}'_n)\left/
      \sqrt{\left(\sum_n(h_n|h_n)\right)\left(\sum_n({\bar h}'_n|{\bar h}'_n)\right)}\right.\, .
\end{equation}
In the actual computation, we truncated the echoes at 30 times. 

\begin{figure}[!t]  
\begin{center}
\includegraphics[bb=0 0 750 511, width=0.48\textwidth,clip=true]{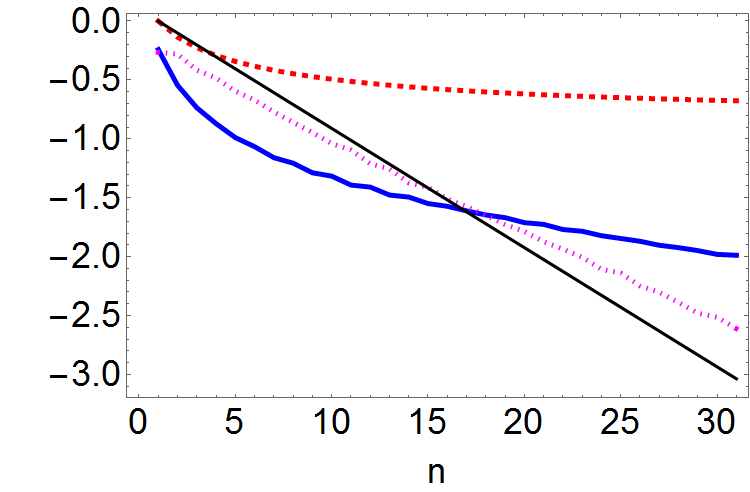}
~~~
\includegraphics[bb=0 0 750 511, width=0.46\textwidth,clip=true]{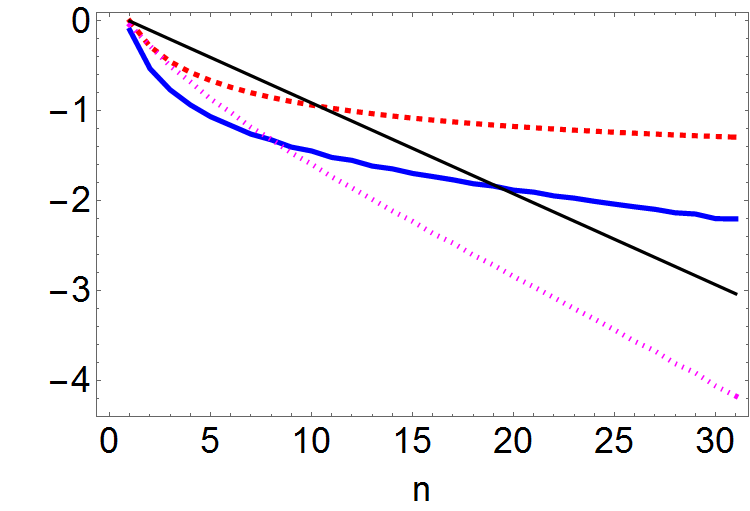}
\end{center}
 \caption{$\log_{10} A_n$ (red dashed curve), $\log_{10} B_n$ (blue thick curve) 
and $\log_{10} C_n$ (magenta dotted curve), as well as
the decline rate $\log_{10} (0.89)^{2n}$ suggested by the analysis 
in Ref.~\cite{Abedi:2016hgu} (black line). The left panel 
is for $\alpha=0.5$ while the right panel is for $\alpha=1.0$.}
\label{Fig:ABC}
\end{figure}  

In Fig.~\ref{Fig:ABC},
we give plots of $A_n$ (red dashed curve), $B_n$ (blue thick curve) 
and $C_n$ (magenta dotted curve) as 
well as the decline rate $(0.89)^{2n}$ (black line)
suggested by the analysis in Ref.~\cite{Abedi:2016hgu}.
For the left and right panels $\alpha$ in $h$ is set to $0.5$ and $1.0$, respectively. 
The decline rate suggested by the slope of $B_n$ 
is shallower than $(0.89)^{2n}$, while that of $C_n$ is slightly shallower 
for $\alpha=0.5$ but slightly steeper for $\alpha=1.0$.   
In both cases, $C_n$ is larger than $B_n$ for small $n$ because
the template with the frequency-dependent reflection rate
in the calculation of $B_n$ gives
larger signal-to-noise ratio from the contribution at larger $n$.

In the above setup we obtained 
\begin{eqnarray}
\rho_B=0.493\,, \qquad
\rho_C=0.443\,. \qquad (\mbox{for}~ \alpha=0.5)\cr
\rho_B=0.729\,, \qquad
\rho_C=0.662\,. \qquad (\mbox{for}~ \alpha=1.0)   
\end{eqnarray}
The value of $\gamma$ that maximizes $\rho_C$ is $\gamma=0.89$ for $\alpha=0.5$,
which is reduced to 0.81 for $\alpha=1.0$.   
Although the difference in the signal-to-noise ratio 
between $\rho_B$ and $\rho_C$ is small, there is a chance that the 
signal becomes slightly more significant (typically about 10\%) 
without increasing the number of parameters
by using the remnant BH parameters derived from data analysis
of the usual BBH merger
if we use the templates that take into account the reflection rate. 
In that case, the damping of the amplitude of echoes 
is expected to be even slower.

\section{Summary}\label{sec:sum}

In this paper, we have proposed a possible improvement of
the GW template for BH echoes,
motivated by the solution of the perturbation equation with
a completely reflecting boundary placed near the horizon.
Although the nature of the boundary is unknown,
it seems natural to assume that there is no complicated frequency
dependence of the phase shift within the narrow frequency
band relevant for BH echoes. Under this
assumption, we can give a template which properly takes into
account the reflection rate at the angular momentum barrier,
without increasing the number of tunable free parameters.
The proposed template for echoes is given in a simple analytic
form, once the waveform of the incident ingoing wave is provided.

\section*{Acknowledgments}

~~~This work was supported by MEXT Grant-in-Aid for Scientific Research
on Innovative Areas,
``New Developments in Astrophysics Through Multi-Messenger Observations
of Gravitational Wave Sources,'' Nos.~24103001 (TT) and~24103006 (HN, TT),
by the Grant-in-Aid from the Ministry of Education, Culture, Sports,
Science and Technology (MEXT) of Japan No.~15H02087 (TT),
and by JSPS Grants-in-Aid for Scientific Research (C), 
No.~16K05347 (HN) and No.~16K05356 (NS).


\end{document}